\documentclass[%
 reprint,
 superscriptaddress,
 amsmath,amssymb,
 aps,
 prl
]{revtex4-1}

\usepackage{amsmath}
\usepackage{amsthm}
\usepackage[colorlinks=true,urlcolor=blue,citecolor=blue,linkcolor=blue]{hyperref}
\usepackage[shortlabels]{enumitem}
\usepackage{amsfonts}
\usepackage{graphicx}
\usepackage{bbold}
\usepackage{stmaryrd}
\usepackage{color}
\usepackage{hyperref}
\usepackage{verbatim}
\usepackage{cases}
\usepackage{amsfonts}
\usepackage{amssymb}
\usepackage[]{mathrsfs}
\usepackage{xcolor}
\usepackage{epsfig}
\usepackage{bm}
\usepackage{setspace}
\usepackage{enumerate}
\usepackage{dcolumn}
\usepackage{bm}
\usepackage{enumitem}
\usepackage{soul}

\usepackage{graphicx}
\usepackage{dcolumn}
\usepackage{bm}
\usepackage[FIGTOPCAP,small]{subfigure}
\usepackage{color}

\newcommand{\ket}[1]{\left\vert #1 \right>}
\newcommand{\pare}[1]{\left( #1 \right)}
\newcommand{\abs}[1]{\left\vert #1 \right\vert}
\newcommand{\cor}[1]{\left[ #1 \right]}

\newcommand{\ave}[1]{\left\langle #1 \right\rangle}

\begin{document}

\preprint{APS/123-QED}

\title{Multiphoton Quantum-State Engineering using Conditional Measurements}

\author{Omar S. Maga\~{n}a-Loaiza}
\email{maganaloaiza@lsu.edu}
\affiliation{Department of Physics and Astronomy, Louisiana State University,
Baton Rouge, Louisiana 70803, USA}
\affiliation{National Institute of Standards and Technology, 325 Broadway, Boulder Colorado 80305, USA}
\author{Roberto de J. Le\'on-Montiel}
\affiliation{Instituto de Ciencias Nucleares, Universidad Nacional Aut\'onoma de
M\'exico,\\ Apartado Postal 70-543, 04510 Cd. Mx., M\'exico}
\author{Armando Perez-Leija}
\affiliation{Max-Born-Institut, Max-Born-Stra{\ss}e 2A, 12489 Berlin, Germany}
\affiliation{Humboldt-Universit\"{a}t zu Berlin, Institut f\"{u}r Physik, AG Theoretische Optik Photonik, Newtonstra{\ss}e 15, 12489 Berlin, Germany}
\author{Alfred B.~U'Ren}
\affiliation{Instituto de Ciencias Nucleares, Universidad Nacional Aut\'onoma de
M\'exico,\\ Apartado Postal 70-543, 04510 Cd. Mx., M\'exico}
\author{Chenglong You}
\affiliation{Department of Physics and Astronomy, Louisiana State University,
Baton Rouge, Louisiana 70803, USA}
\author{Kurt Busch}
\affiliation{Max-Born-Institut, Max-Born-Stra{\ss}e 2A, 12489 Berlin, Germany}
\affiliation{Humboldt-Universit\"{a}t zu Berlin, Institut f\"{u}r Physik, AG Theoretische Optik Photonik, Newtonstra{\ss}e 15, 12489 Berlin, Germany}
\author{Adriana E. Lita}
\affiliation{National Institute of Standards and Technology, 325 Broadway, Boulder Colorado 80305, USA}
\author{Sae Woo Nam}
\affiliation{National Institute of Standards and Technology, 325 Broadway, Boulder Colorado 80305, USA}
\author{Richard P. Mirin}
\affiliation{National Institute of Standards and Technology, 325 Broadway, Boulder Colorado 80305, USA}
\author{Thomas Gerrits}
\affiliation{National Institute of Standards and Technology, 325 Broadway, Boulder Colorado 80305, USA}
\date{\today}

\begin{abstract}
The quantum theory of electromagnetic radiation predicts characteristic statistical fluctuations for \textcolor{black}{light sources as diverse as} sunlight, laser radiation and molecule fluorescence. Indeed, these underlying statistical fluctuations of light are associated with the fundamental physical processes behind their generation. In this contribution, we demonstrate that the manipulation of the quantum electromagnetic fluctuations of a pair of vacuum states leads to a novel family of quantum-correlated multiphoton states with tunable mean photon numbers and degree of correlation. Our technique relies on the use of conditional measurements to engineer the excitation mode of the field through the simultaneous subtraction of photons from two-mode squeezed vacuum states. The experimental generation of multiphoton states with quantum correlations by means of photon subtraction unveils novel mechanisms to control fundamental properties of light. As a remarkable example, we demonstrate the engineering of a quantum correlated state of light, with nearly Poissonian photon statistics, that constitutes the first step towards the generation of entangled lasers. Our technique enables a robust protocol to prepare quantum states with multiple photons in high-dimensional spaces \textcolor{black}{and,} as such, it constitutes a novel platform for exploring quantum phenomena in mesoscopic systems.



\end{abstract}

\pacs{Valid PACS appear here}
\maketitle

The identification of the photon as a fundamental carrier of information has triggered a wide variety of research that aims to improve the state of the art of photonic technologies \cite{brien2009}. Along these lines, the field of quantum photonics has focused on exploiting the quantum properties of light to dramatically improve the performance of protocols for communication, metrology, imaging, and information processing \cite{giovannetti2011,aspuru2012,willner2015}. Consequently, the successful implementation of functional quantum photonic technologies hinges on our ability to generate, manipulate, and measure  complex multiphoton states \cite{dellanno2006, pan2012, hadfield2009}. However, the generation of high-dimensional entangled states comprising a large number of photons is nowadays one of the most challenging tasks in quantum optics \cite{huber2010,hiesmayr2016}.

Over the past few years, physicists and engineers have demonstrated the \textcolor{black}{utilization of} multiple degrees of freedom of single photons to perform information processing tasks for a wide range of applications. In this regard, multiple bits of information have successfully been encoded in a single photon by preparing complex superpositions in time, frequency, position, transverse momentum, angular position, orbital angular momentum and path \cite{xie2015,dixon2012,khan2007,kues2017,dada2011, maga2016}. The complexity of such superpositions has led to important improvements in the performance and tolerance of cryptographic protocols, in the estimation of small physical parameters, and in the development of novel schemes for information processing \cite{brien2009,giovannetti2011,aspuru2012,willner2015}. Furthermore, it is now recognized that such protocols can be further improved by incorporating a high number of photons, correlated and entangled \cite{wen2007-1,wen2007-2,tsang2009,wen2010}.
Interestingly, the generation of a large number of entangled photons has been demonstrated through the use of stimulated parametric down-conversion \cite{eisenberg2004}. Furthermore, the preparation of three indistinguishable particles and their interference signatures have been studied in Ref. \cite{menssen2017}. In 2016, four-photon entanglement in the basis of orbital angular momentum was reported \cite{hiesmayr2016}. More recently, boson sampling with up to five particles was demonstrated by means of multiplexing of single-photon sources \cite{wang2016}.
\begin{figure*}[t]
\hbox{\hspace{0mm} \includegraphics[width=17.87cm]{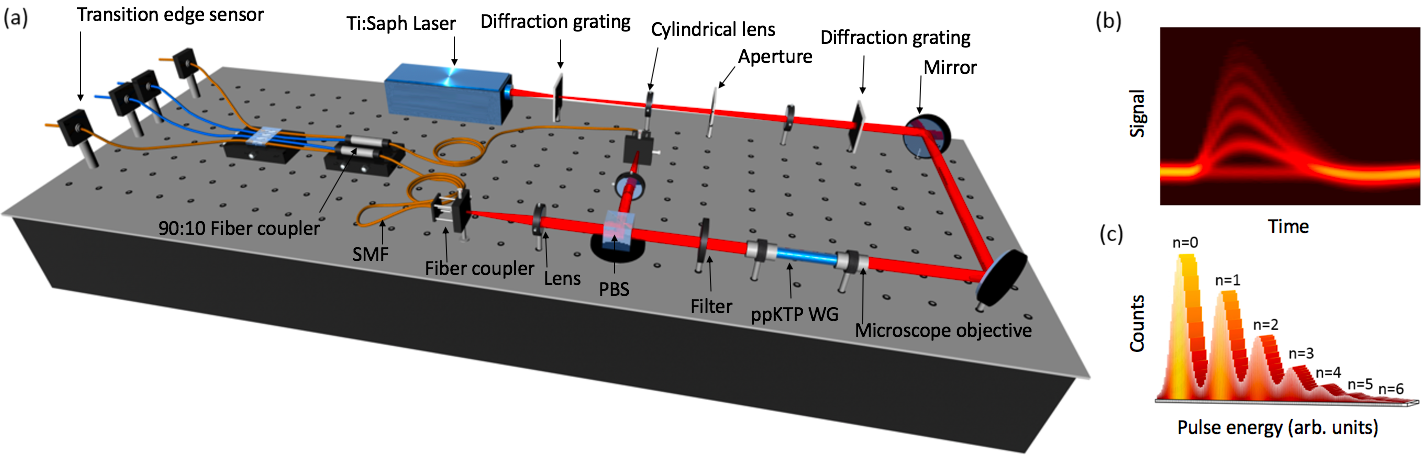}}
\caption{A schematic representation of the experimental setup.
(a)  We generate two-mode squeezed vacuum states (TMSVS) through a type-II parametric down-conversion process. This is achieved by pumping a periodically poled potassium titanyl phosphate waveguide (ppKTP WG) with a femtosecond Ti:Sapphire laser at 785 nm, cavity dumped with a repetition rate of 76 MHz, which are then pulse-picked at a repetition rate of 229.166 kHz. We utilize two diffraction gratings in combination with a 4$f$ optical system to filter the spectral profile of the pump beam. The down-converted photons are passed through a silicon filter, split by a polarizing beam splitter (PBS) and coupled into single mode fibers (SMFs). The simultaneous photon subtraction is performed on the down-converted signal and idler photons by means of two 90:10 fiber couplers that direct photons to four transition edge sensors (TESs) with photon number resolution. The electrical output traces generated by one of our TES are shown in (b). The discrete levels of the signal demonstrate the number resolving capability of our detectors. The traces in (b) are used to obtain the pulse-energy distribution in (c). This information is then used to determine photon number distributions for each of the modes, $n$ represents the number of photons in a given pulse.}
\label{Fig:Setup}
\end{figure*}

Regardless of \textcolor{black}{the} enormous effort \textcolor{black}{directed} thus far \textcolor{black}{at improving} the efficiency of multiphoton sources  \cite{eisaman2011,kwiat1995}, the challenges involved in the generation and manipulation of entanglement and correlations among the photons impose practical limitations to realistic quantum technologies \cite{aspuru2012}. In 1997, Dakna and coworkers introduced the concept of photon subtraction and the possibility of performing quantum state engineering at a new fundamental level in which the quantum vacuum and the excitation mode of the fields are manipulated \cite{dakna1997}. This seminal work triggered a wide variety of fundamental and applied research, including new fundamental tests of quantum mechanics, engineering of macroscopic states of light, and the development of novel schemes for quantum metrology \cite{ourjoumtsev2006,parigi2007,hashemi2017}. In particular, regarding quantum-state engineering, Takahashi $et$ $al.$ \cite{takahashi2010} demonstrated distillation of entanglement from a single-mode squeezed state and four years later Kurochkin and colleagues demonstrated entanglement distillation by subtracting photons from standard two-mode squeezed vacuum states (TMSVS) \cite{kurochkin2014}. More recently, in the same context, Carranza and Gerry predicted the possibility of generating mesoscopic states of light with tunable average photon numbers by subtracting an equal number of photons from each mode of TMSVS \cite{carranza2012}.

Standard sources of TMSVS are based on the process of spontaneous-parametric-down-conversion (SPDC) and, under ideal conditions, they produce infinite entangled superpositions of photons as described by the expression $\ket{z} = \sqrt{1-\abs{z}^{2}}\sum_{n=0}^{\infty}z^{n}\ket{n}_{s}\ket{n}_{i}$, where $n$ denotes the number of photons in the signal ($s$) and idler ($i$) modes, and $z$ represents the so-called squeezing parameter, which depends on the material properties. Under realistic conditions, however, SPDC sources operate at very low brightness and the probability of generating more than one photon per mode is extremely low. This is in fact the main limitation for the efficient generation of photon-subtracted TMSVS based on standard SPDC sources. In this work, we circumvent this challenge by utilizing a bright SPDC source \cite{harder2016} in combination with photon-number-resolving (PNR) detectors \cite{lita2008}, to demonstrate the generation of a novel family of  correlated photon-subtracted TMSVS with a broad range of mean photon numbers and degrees of correlation. This is achieved by engineering quantum statistical fluctuations of light via conditional measurements. Moreover, our observations reveal that the generated states are composed of two multiphoton wavepackets which are highly correlated and exhibit nearly Poissonian statistics. Remarkably, this achievement constitutes an important step towards the development of entangled laser-like systems.
Since this family of states naturally inhabits high-dimensional Hilbert spaces, we foresee their applicability to enrich protocols relying on multiphoton interference such as high-sensitivity interferometry, quantum imaging, and sub-Rayleigh lithography \cite{wen2007-1,wen2007-2,tsang2009,wen2010}.

%
\section{Generation of photon-subtracted TMSVS}
Photon-subtracted TMSVS containing the same number of photons in each mode exhibit the highest degree of correlation \cite{carranza2012,barnett1989}. Consequently, we restrict ourselves to the case of symmetric photon-subtracted TMSVS. Mathematically, the subtraction of $l$-photons from ideal TMSVS is obtained by applying $l$-times the bosonic annihilation operators $\hat{a}_{s,i}$ to the TMSVS $\ket{z}$ yielding $\ket{z,-l}=(n_s n_i)^{-1/2}\ket{z,-l}=\hat{a}_{s}^{l}\hat{a}_{i}^{l}\ket{z}=\sum_{j=l}^{\infty}B_{j}^{\pare{l}}\ket{j-l}_{s}\ket{j-l}_{i}$, where $B_{j}^{\pare{l}}=\mathcal{N}^{-1/2}\cor{j!/\pare{j-l}!}z^{j}$, and the normalization constant $\mathcal{N}=\sum_{n=0}^{\infty}\abs{z}^{2j}\cor{j!/\pare{j-l}!}^2$. \textcolor{black}{Note that} by construction photon-subtracted TMSVS constitute a family of entangled states characterized by the correlation function $B_{j}^{\pare{l}}$, which depends on the number of subtracted photons $l$ and the squeezing parameter $z$. The above expression is constructed for perfect TMSVS, close to unity beam splitter transmission and independent of losses. Below, we incorporate the optical losses due to the beam splitter transmission into a single efficiency value, a good approximation to our experiment. This implies that the degree of nonclassical correlations in photon-subtracted TMSVS can be readily controlled by tuning these parameters.

For our experiments we utilize a Ti:Sapphire laser and a spectral \textcolor{black}{filter}, composed \textcolor{black}{of} two gratings and a 4$f$ optical system, to produce 1 ps optical pulses that are used to pump a periodically poled potassium titanyl phosphate (ppKTP) waveguide with a length $L$ of $8$ mm, a nominal poling period of 46 $\mu$m and a width of 2 $\mu$m, see Fig.~(\ref{Fig:Setup}~a). This configuration allows us to produce bright TMSVS at a wavelength of 1570 nm by means of type-II parametric down-conversion process  \cite{harder2016}. For this bright SPDC source we have the squeezing parameter given as $z=\text{tanh}\pare{r}$, with $r=\chi_{\text{eff}}\omega_{p} L\sqrt{I_{p}}/\pare{2n_{0}c}$, where $n_{0}$ is the refractive index and $\chi_{\text{eff}}$ is the effective nonlinearity of the waveguide, $\omega_{p}$ is the frequency and $I_{p}$ the intensity of the pump field.\\
\begin{figure*}[t!]
\centering
\includegraphics[width=\textwidth]{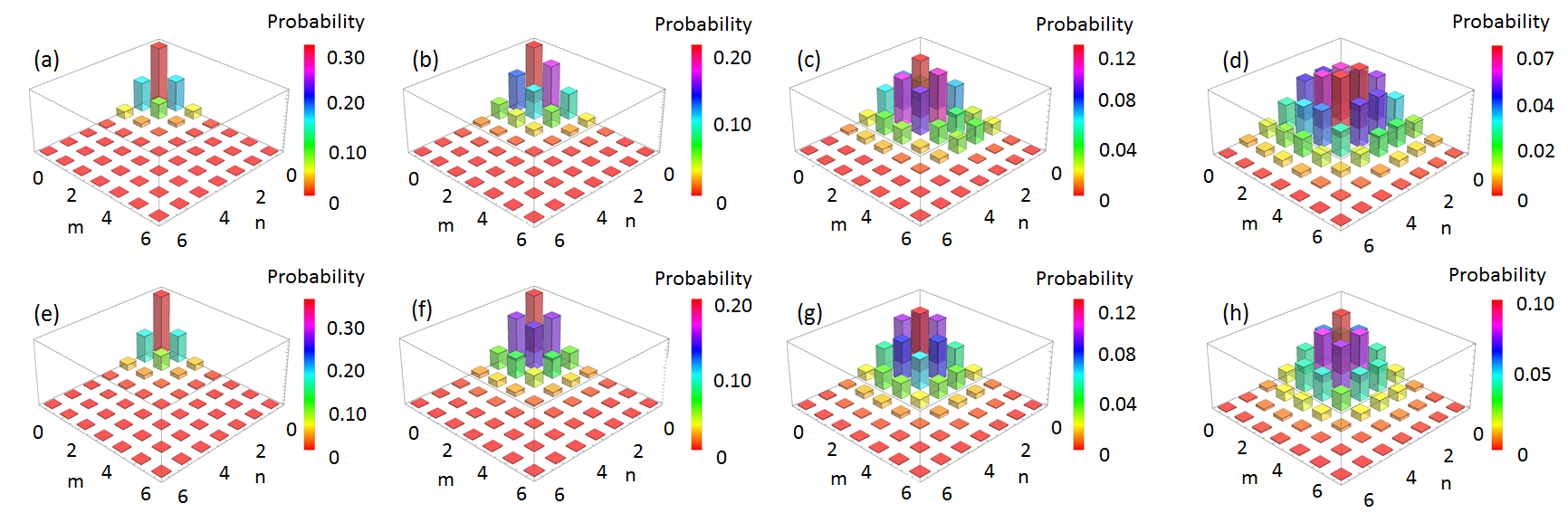}
\caption{Joint photon number distributions for  photon-subtracted TMSVS. (a to d) The first row shows the experimental joint photon number distributions for the simultaneous subtraction of zero ($l=0$), one ($l=1$), two ($l=2$) and three ($l=3$) photons. (e to h) The bottom row shows the theoretical predictions obtained by evaluating Eq. (\ref{Eq:photon_distribution_with_losses}). We assume an overall system efficiency of approximately $16.25\%$ and a squeezing parameter $z$ of $0.66$.}
\label{fig:Distribution3D}
\end{figure*}
The simultaneous subtraction of $l$-photons is achieved by routing \textcolor{black}{each of} the signal and idler modes into \textcolor{black}{a} 90:10 fiber coupler. Each coupler subtracts photons by directing approximately 10$\%$ of the injected photons to a transition edge sensor (TES) that acts as a photon number resolving detector. The typical electrical output traces generated by a TES are shown in Fig.~(\ref{Fig:Setup}~b). The measured pulse energy due to the absorbed photons, shown in Fig.~(\ref{Fig:Setup}~c), clearly demonstrates the photon-number resolution of the TESs \cite{lita2008}. A series of Gaussian fits to the pulse-energy distributions are then used to estimate photon number probabilities. Crucially, the photon events detected by the central TESs \textcolor{black}{permit} conditional measurements on the external TESs that register the remaining 90$\%$ of the input photons. We then monitor the correlations between the photon-subtracted states.\\
\begin{figure}[b!]
\centering
\includegraphics[width=8cm]{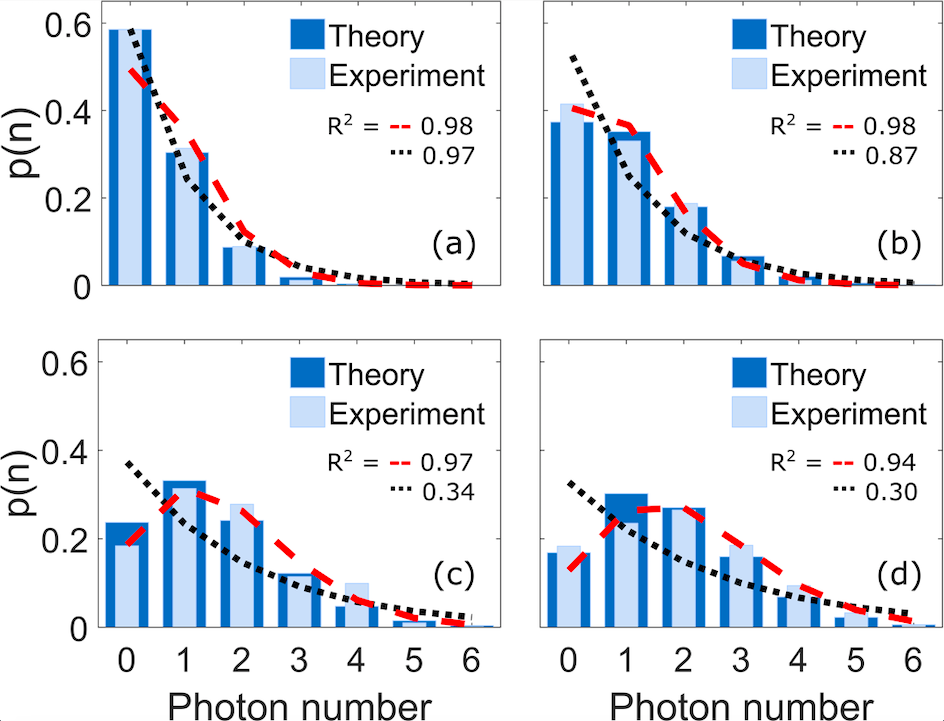}
\caption{Marginal photon number distributions for photon-subtracted TMSVS with $l$. (a) $l=0$, (b) $l=1$, (c) $l=2$, and (d) $l=3$. The overall system's efficiency is approximately 16.25$\%$ and the best-fit squeezing parameter is obtained to be 0.66. The red-dashed and black-dotted curves represent Poissonian and thermal photon distributions, respectively. These curves are obtained from a fit to the experimental data with mean photon numbers equal to (a) $\ave{n} = 0.7$,  (b) $\ave{n} = 0.9$, (c) $\ave{n} = 1.68$, and (d) $\ave{n} = 2.05$. \textcolor{black}{The coefficients of determination ($\text{R}^2$) for the distribution fittings are shown in each panel.}}
\label{Fig:Marginal}
\end{figure}
In Fig.~(\ref{fig:Distribution3D}) we present the measured joint photon number distributions (top row) for the first generation of quantum-correlated photon-subtracted TMSVS inhabiting in a multi-dimensional space for different number of subtracted photons, (a) $l=0$, (b) $l=1$, (c) $l=2$, and (d) $l=3$.
Importantly, we observe that photon subtraction generates a family of quantum states, whose marginal photon statistics \textcolor{black}{are} modified from thermal to nearly-Poissonian, with increasingly larger mean photon numbers, see Fig.~(\ref{Fig:Marginal}). These effects become more evident for the case of $l=3$ and $z=0.66$, see Fig.~(\ref{Fig:Marginal}~d). In that particular case, we clearly observe that the marginal photon number distributions for the signal and idler modes, strongly resemble the photon number distribution that characterizes a laser (red dashed curves in Fig. 3) with a mean photon number of $\ave{n}=2.05$ and not that of a thermal distribution as in conventional SPDC sources (black dotted curve in Fig.3). In other words, the generated light field is composed of two wavepackets exhibiting nearly Poissonian statistics. In that sense, the generated states share some similarities with a system of entangled lasers \cite{carranza2012}.

In order to characterize the generated states we use a theoretical model that assumes perfect TMSVs and photon detectors with \textcolor{black}{non-unit} detection efficiencies and contributions from dark counts. These are reasonable assumptions for sources of this kind as described in \cite{harder2016, sergey2017}.  This models allows the computation of the realistic probability of simultaneously detecting $m$ photons in the signal and $n$ photons in the idler mode (see Supplementary Materials)
\small
\begin{equation}\label{Eq:photon_distribution_with_losses}
\begin{split}
p(n,m &)  =  \frac{1}{n!m!}\sum_{j,k=l}^{\infty} \frac{B_{j}^{\pare{l}}B_{k}^{\pare{l}*}}{\pare{j-l}!\pare{k-l}!} \\
&  \times \partial_{\alpha}^{j-l}\partial_{\alpha^{*}}^{k-l}\left\lbrace \pare{\eta\alpha^{*}\alpha + \nu}^{n}e^{-\cor{\pare{\eta -1}\alpha^{*}\alpha + \nu}}\right\rbrace\Big\vert_{\alpha,\alpha^{*}=0} \\
& \times \partial_{\beta}^{j-l}\partial_{\beta^{*}}^{k-l}\left\lbrace \pare{\eta\beta^{*}\beta + \nu}^{m}e^{-\cor{\pare{\eta -1}\beta^{*}\beta + \nu}}\right\rbrace\Big\vert_{\beta,\beta^{*}=0}.
\end{split}
\end{equation}
\normalsize
Here, the parameters $\eta < 1$ and $\nu> 0$ represent the overall detector quantum efficiency\textcolor{black}{, including the effect of all losses in the setup,} and the dark counts, respectively \cite{sperling2012}. Our model considers losses only after beam splitters and assumes beam splitters with almost perfect transmissivity. \textcolor{black}{Note that} the complex parameters $\alpha$ and $\beta$ arise from the Glauber-Sudarshan $P$-function corresponding to the photon-subtracted TMSVS \cite{sudarshan1963,glauber1963,mehta1967,book_knight}. It is worth remarking that the theoretical model used to obtain Eq.~(\ref{Eq:photon_distribution_with_losses}) can be readily extended to describe an asymmetric subtraction of photons from TMSVS, as well as photon subtraction from any two-mode quantum field (see Supplementary Materials).
\begin{figure*}[t]
\hbox{\hspace{7.8mm} \includegraphics[width=15.9cm]{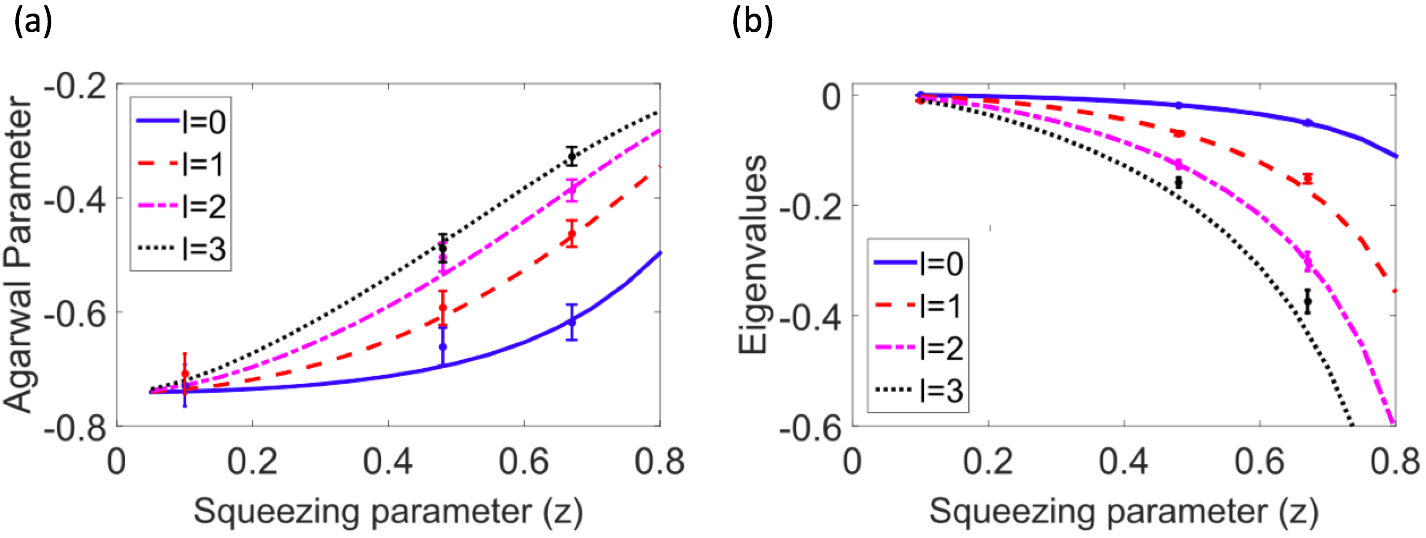}}
\caption{Agarwal parameter and eigenvalues for the second-order matrix of moments $M^{(2,2)}$ for the generated photon-subtracted TMSVS as a function of the squeezing parameter. The Agarwal parameter in (a) certifies the nonclassical nature of $P$-function corresponding to the generated photon-subtracted states for different number of subtracted photons $l$. The eigenvalues in (b) demonstrates the possibility of tuning the degree of quantum correlations by controlling the squeezing parameter of the source and the number of subtracted photons. The curves represent our theoretical predictions for $l=0$ (blue solid line), $l=1$ (red dashed line), $l=2$ (pink dash-dotted line), and $l=3$ (black dotted line). The points with error bars correspond to the Agarwal parameter and eigenvalues extracted from the experimentally measured photon number distributions.}
\label{Fig:Entropy}
\end{figure*}
Our theoretical results, shown in Fig.~(\ref{fig:Distribution3D}) (bottom row) are obtained by evaluating Eq.~\eqref{Eq:photon_distribution_with_losses}, with the best fit corresponding to an overall system efficiency of $16.25\%$. We compare this to our estimation of the overall efficiency of our setup. The source efficiency, beam splitter transmission and detection efficiency are estimated to be 90 $\%$, 90 $\%$ and 80 $\%$, leaving a fiber coupling efficiency of about 25 $\%$.
\section{Non-classicality test of photon-subtracted TMSVS}
A natural metric to investigate the non-classicality of the photon-subtracted states is through applying the Cauchy-Schwarz inequality $\langle (\hat{a}_{s}^{\dagger})^2 \hat{a}_{s}^2 \rangle\langle (\hat{a}_{i}^{\dagger})^2 \hat{a}_{i}^2 \rangle\geq |\langle \hat{a}_{s}^{\dagger}\hat{a}_{s}\hat{a}_{i}^{\dagger}\hat{a}_{i} \rangle|^{2}$, which is fulfilled by any classical two-mode light field. That is, the inequality is obeyed if and only if the corresponding P-function, $P(\alpha,\beta)$, is a classical probability distribution.
Accordingly, any violation of this inequality unequivocally indicates the nonclassical nature of any bimodal state, or equivalently that $P(\alpha,\beta)$ is non-classical \cite{Agarwal1988}. As a measure of this violation, we consider the Agarwal parameter
\begin{align}
I=\frac{\sqrt{\langle (\hat{a}_{s}^{\dagger})^2 \hat{a}_{s}^2 \rangle\langle (\hat{a}_{i}^{\dagger})^2 \hat{a}_{i}^2 \rangle}}{\langle \hat{a}_{s}^{\dagger}\hat{a}_{s}\hat{a}_{i}^{\dagger}\hat{a}_{i} \rangle}-1,
\label{eq:Agarwal}
\end{align}
which is negative if the inequality is violated \cite{Agarwal1988}. In Fig.~(\ref{Fig:Entropy}) we present the experimental Agarwal parameter $I$ as a function of the squeezing parameter $z$ for the photon-subtracted states $\ket{z,-l}$ with $l=0, 1, 2,$ and $3$. The theoretical curves are computed using Eq.~\eqref{eq:Agarwal} and the photon-number distributions obtained for our realistic experimental conditions through Eq.~\eqref{Eq:photon_distribution_with_losses}.
For small values of squeezing $(z\rightarrow0)$ we observe the greatest degree of violation of the inequality. Conversely, by increasing the \textcolor{black}{degree of} squeezing $(z\rightarrow1)$ $I$ approaches zero.
At this point it is worth emphasizing that a product of coherent states, \textcolor{black}{which represents the classical-quantum boundary}, gives an Agarwal parameter $I = 0$. Hence, it is tempting to think that as we subtract \textcolor{black}{an increasing number of photons} and $(z\rightarrow1)$, the state becomes a product of coherent states. However, as we show below, this is not the case, since at this particular regime the generated states become highly correlated. \\
To analyze the nature of the correlations between the generated photon wavepackets we explore the two-dimensional characteristic function $\sum_{m,n=0}^{\infty}p(m,n)x^{m}y^{n}$, where $p(m,n)$ is the probability of simultaneously detecting $m$ photons in the signal and $n$ photons in the idler modes.
Then, by defining the operators $\hat{m}_{s(i)}=(\eta \hat{n}_{s(i)}+\nu)$ we obtain the joint factorial moments
\small
\begin{align}
\begin{split}
\left\langle: \hat{m}_{s}^{u}\otimes \hat{m}_{i}^{v}:\right\rangle=\sum_{m,n=0}^{\infty}p(m,n)m(m-1)...(m-u+1)\\\times
n(n-1)...(n-v+1).
\end{split}
\label{eq:JM}
\end{align}
\normalsize
Using Eq.~\eqref{eq:JM}, it is possible to construct a matrix of moments of any order. However, it is sufficient to explore the behavior of the second order matrix of moments to demonstrate the nonclassical nature of the correlations between modes
\begin{align}
M^{(2,2)}=
\begin{pmatrix}
\langle: \hat{m}_{s}^{0} \hat{m}_{i}^{0}:\rangle&  \langle: \hat{m}_{s}^{1} \hat{m}_{i}^{0}:\rangle & \langle: \hat{m}_{s}^{0} \hat{m}_{i}^{1}:\rangle\\
\langle: \hat{m}_{s}^{1} \hat{m}_{i}^{0}:\rangle&  \langle: \hat{m}_{s}^{2} \hat{m}_{i}^{0}:\rangle &  \langle: \hat{m}_{s}^{1} \hat{m}_{i}^{1}:\rangle\\
\langle: \hat{m}_{s}^{0} \hat{m}_{i}^{1}:\rangle&  \langle: \hat{m}_{s}^{1} \hat{m}_{i}^{1}:\rangle &  \langle: \hat{m}_{s}^{0} \hat{m}_{i}^{2}:\rangle
  \end{pmatrix}.
\end{align}
Crucially, the determinant of $M^{(2,2)}$ yields information about the mean values, the variances, and the covariance of the joint-photon statistics $\text{det}(M^{(2,2)}) = \langle: (\Delta \hat{m}_{s})^{2}:\rangle\langle: (\Delta \hat{m}_{i})^{2}:\rangle-\langle: \Delta \hat{m}_{s}\Delta \hat{m}_{i}:\rangle^{2}$.
For any two-mode classical field the covariance square $\langle:\Delta\hat{m}_{s}\Delta\hat{m}_{i}:\rangle^{2}$ is always greater than the product of the marginal variances $\langle:(\Delta \hat{m}_{s})^{2}:\rangle\langle:(\Delta\hat{m}_{i})^{2}:\rangle$ (see Supplementary Materials). Thus, \textcolor{black}{in order} to show the non-classicality of the correlations among the signal and idler modes, it is sufficient to show the contrary, that is, $\text{det}(M^{(2,2)})<0$.\\
In Table~(\ref{tab:T1}), we present the experimental $\text{det}(M^{(2,2)})$ for $l=0, 1, 2, 3$ and for three different values of $z$. Quite remarkably, as we subtract more photons, the corresponding determinant becomes more negative. This implies that by subtracting a higher number of photons the covariance square becomes much larger than the product of marginal variances, thus indicating an increasingly larger degree of correlation between modes.\\
\small
\begin{table}
  \centering
\begin{tabular}{ |c |c |c |c |c |c |c |c| }
 \hline
 z & l=0 &  l=1 & l=2 & l=3 \\
  \hline
0.1 & $-4.06\text{x}10^{-7}$  & $-4.82\text{x}10^{-6}$& $-3.05\text{x}10^{-5*}$ & $-8.79\text{x}10^{-5*}$ \\
0.48 & $-0.55\text{x}10^{-3}$ & $-6.09\text{x}10^{-3}$ & $-19.54\text{x}10^{-3}$ & $-15.22\text{x}10^{-3}$ \\
0.67 & $-6.91\text{x}10^{-3}$ & $-62.7\text{x}10^{-3}$ & $-202.4\text{x}10^{-3}$ & $-332.2\text{x}10^{-3}$\\
 \hline
\end{tabular}
\caption{Determinant of the second-order matrix of moments $M$. Notice data with $(*)$ indicate a theoretical value predicted by equations 3 and 4 \textcolor{black}{using the fit values for $z$, $\eta$, $\nu$.}}
\label{tab:T1}
\end{table}
\normalsize
A further criterion to demonstrate the non-classicality of the correlations is to probe the non-negativity of $M^{(2,2)}$. That is, since the matrices of moments are non-negative (positive definite) for classical states, any violation of the non-negativity implies genuine non-classical second order correlations for the generated photon-subtracted TMSVS (see Supplementary Materials). In order to demonstrate the non-negativity of $M$, we use its normalized eigenvectors $\mathbf{V}$ and the corresponding eigenvalues $\lambda$.
Since $\mathbf{V}^{\dagger}M^{(2,2)}\mathbf{V}=\lambda$, if all the eigenvalues of $M^{(2,2)}$ are non-negative, it holds that $M^{(2,2)}\geq0$. On the contrary, if at least one eigenvalue is negative we have a violation of the non-negativity of $M^{(2,2)}$, that is, $M^{(2,2)}\ngeq0$. Here, we have computed the minimal eigenvalues of $M^{(2,2)}$ and the results are shown in Fig.~(\ref{Fig:Entropy} b). Clearly, all the eigenvalues are negative and we can conclude that $M^{(2,2)}\ngeq0$, as a result, the correlations between the idler and signal modes are nonclassical.\\
\section{Conclusions}
The possibility of controlling multiphoton states in high-dimensional Hilbert spaces paves the way towards novel quantum protocols in the mesoscopic regime, being of particular relevance for quantum simulators \cite{konrad2018}, quantum optical interferometry \cite{dowling2008}, and quantum networks \cite{ourjoumtsev2009,sanders2012,israel2018}. More importantly, the ability of conditioning the mean photon number of photon-subtracted states enables the possibility of using our setup to perform experiments where heralding of multiple photons is required, a task that remains challenging \cite{harder2016}. Along these lines, in this work we have demonstrated the first experimental generation of a novel family of quantum-correlated photon-subtracted states with tunable mean photon numbers and degree of quantum correlations. 
 We certified the quantum nature of the correlations in the generated photon-subtracted states through the use of the Agarwal parameter. Furthermore, the control of the degree of quantum correlations by means of photon subtraction was quantified. Given the enormous interest in high-dimensional quantum photonics with multiple particles, we anticipate the use of our technique for testing fundamental multiphoton physics \cite{pan2012}, and for the development of novel schemes for quantum metrology and information processing \cite{brien2009, giovannetti2011, aspuru2012, boto2000}.


\section{Acknowledgements}

We acknowledge J. P. Dowling, S. Glancy and D. Reddy for helpful discussions. We thank N. Montaut for useful advices regarding the alignment of the waveguide source.

Contribution of NIST, an agency of the U.S. government, not subject to copyright

\section{Methods}

\textbf{Detectors fabrication}  The superconducting transition edge sensor (TES) is a highly-sensitive microcalorimeter that measures the amount of heat absorbed in the form of photons.  By exploiting the sharp superconducting-to-normal resistive transition in a superconductor as an extremely sensitive thermometer, we can measure the change in resistance from absorbing one or more photons.  Absorption of two photons causes a temperature rise that is ideally twice that of a single photon.  As a result, the TES generates an output signal that is proportional to the cumulative energy in an absorption event.  This proportional pulse-height enables the determination of the energy absorbed by the TES and consequently the direct conversion of sensor pulse-height into photon number.  

We utilize Tungsten (W) as the superconductor material for the fabrication of the TESs used in this work. The device size is 20 $\times$ 20 $\mu$m$^2$, with a superconducting transition temperature of $\sim$150 mK.  The W film is embedded in an optical stack designed to maximize photon absorption at the target wavelength, in this case 1550 nm.  The optical stack consists of a highly reflective bottom mirror (electron-beam evaporated Ag film $\sim$80 nm), a dielectric spacer (physical vapor deposition (PVD) grown layer of SiO2  $\sim$232 nm thick plus a layer of direct current (DC)-sputtered amorphous-Si (a-Si) $\sim$20 nm thick), the active detector layer (DC-sputtered high-purity W, 20 nm thick) and an anti-reflection (AR) coating (DC-sputtered a-Si $\sim$56 nm and a PVD-grown layer of SiO2 $\sim$183 nm). Our fiber-packaging scheme ensures sub-micron alignment of the device active area to the fiber core of SMF-28 standard telecom fiber.  The optimized optical stack as well as the packaging scheme ensures detection efficiency $>$ 95$\%$ of photons transmitted through the fiber. 

\textbf{Detectors operation} TES are required to operate at low temperatures, below their transition temperature. Therefore, elaborate cooling systems are required to operate these devices. We operate the TESs in a commercial dilution refrigerator at a temperature around 30 mK. To operate the TESs, we voltage-bias the TESs which allows stable operation within the transition region [M1]. The TES readout is typically accomplished by use of superconducting quantum interference devices (SQUIDs) [M2], which serve as low noise amplifiers of the current flowing through the TES. The TES/SQUID responses due to different optical energies, i.e. photon numbers, are further amplified by room-temperature electronics and digitized. To maximize the signal to noise ratio, we perform post-processing of the output waveforms. One fast, reliable method is optimum filtering using a Wiener filter [M3]. This method yields a projection of each waveform onto a known template, resulting in an outcome that is approximately proportional to the absorbed energy, hence the number of photons for a given absorption event. 

\noindent [M1] Irwin, K.D., An application of electrothermal feedback for high resolution cryogenic particle detection. App. Phys. Lett., 66(15), 1998-2000 (1995).

\noindent [M2] Jaklevic, R., et al., Quantum Interference Effects in Josephson Tunneling. Phys. Rev. Lett., 12(7), 159-160  (1964).

\noindent [M3] Fixsen, D.J., et al., Pulse estimation in nonlinear detectors with nonstationary noise. Nucl. Instrum. Methods Phys. Res. A, 520(1–3), 555-558 (2004).

\end{document}


\begin{flushleft}
\singlespacing{\Large{\textbf{Supplementary Materials:}\vspace{2mm}\\
\Large{\textbf{Multiphoton Quantum-State Engineering using Conditional Measurements}}}}
\end{flushleft}
\vspace{0.5cm}

\section*{\large{1. Asymmetric subtraction of photons}}
In general, the probability of detecting $m$ photons in a single-mode field with a photon number resolving detector is computed by the expectation value of the operator $\ket{m}\bra{m}$ in normal order, $p(m) = \left\langle:\frac{(\eta\hat{n}+\nu)^{m}}{m!}e^{-(\eta\hat{n}+\nu)}:\right\rangle$, where $\hat{n}$ is the number operator and $\eta< 1$ and $\nu> 0$ are the finite quantum efficiency of the detector and the corresponding noise (dark) counts. For a bimodal field we have the joint probability given as
\begin{align}
p(m,n) = \left\langle:\frac{(\eta\hat{n}_{s}+\nu)^{m}}{m!}e^{-(\eta\hat{n}_{s}+\nu)} \otimes \frac{(\eta\hat{n}_{i}+\nu)^{n}}{n!}e^{-(\eta\hat{n}_{i}+\nu)}:\right\rangle,
\label{eq:Pmn}
\end{align}
where the subscripts $s$ and $i$ are the mode labels for signal and idler. To derive $p(m,n)$ for the case where $l_{1}$ and $l_{2}$ photons have been subtracted from each mode of a TMSVS, we first express $p(m,n)$ in terms of the two-mde Glauber-Sudarshan function $P$-function \cite{sudarshan1963,glauber1963,sperling2012}
\begin{equation}
\begin{split}
p\pare{n,m}=\frac{1}{n!m!} & \int d^{2}\alpha d^{2}\beta P\pare{\alpha,\beta} e^{-\pare{\eta\abs{\alpha}^2 + \nu}} e^{-\pare{\eta\abs{\beta}^2 + \nu}}\cor{\eta\abs{\alpha}^{2} + \nu}^{n} \cor{\eta\abs{\beta}^{2} + \nu}^{m}.
\end{split}
\label{Eq:Photon_dist}
\end{equation}
The asymmetric subtraction of photons from TMSVS is obtained by considering the most general case in which $l_1$ photons are subtracted from the signal mode, whereas $l_2$ photons are removed from the idler mode. In this scenario, the resulting state can be represented as
\begin{equation}
\ket{\psi}_{l_1,l_2} = \mathcal{N}\sqrt{1-\abs{z}^2}\sum_{j=0}^{\infty}z^{j}\hat{a}_{s}^{l_1}\hat{a}_{s}^{l_2}\ket{j}_s\ket{j}_i,
\end{equation}
where $\mathcal{N}$ is a normalization constant.
It is straightforward to show that after applying the annihilation operators on each of the modes, Eq. (S.2) can be written as
\begin{equation}
\ket{\psi}_{l_1,l_1+l} = \sum_{j=l_1}^{\infty}B_{j}^{\pare{l_1}}\ket{j-l_1}_s\ket{j-l_1-l}_i,
\end{equation}
with
\begin{equation}
B_{j}^{\pare{l_1}} = \mathcal{N}_{-l_1}\pare{\frac{j!}{\sqrt{\pare{j-l_1}!\pare{j-l_1-l}!}}}z^{j},
\end{equation}
and
\begin{equation}
\mathcal{N}_{-l_1} = \cor{\sum_{j=l_1}^{\infty}\abs{z}^{2j}\pare{\frac{j!}{\pare{j-l_1}!}}\pare{\frac{j!}{\pare{j-l_1-l}!}}}^{-1/2}.
\end{equation}
Notice that, without loss of generality, we have $l_2=l_1+l$. By using the state in Eq. (S.3) we can write the density matrix of the asymmetric photon-subtracted state as
\begin{equation}
\hat{\rho}_{l_1,l_1+l} = \sum_{j,k=l_1}^{\infty}B_{j}^{\pare{l_1}}B_{k}^{*\pare{l_1}}\ket{j-l_1}_s\ket{j-l_1-l}_i\bra{k-l_1-l}_i\bra{k-l_1}_s.
\end{equation}
Interestingly, the form of the density matrix allows us to write its corresponding $P$ function as \cite{mehta1967,book_knight}
\begin{eqnarray}
P\pare{\alpha,\beta} &=& \sum_{j,k=l_1}^{\infty}\frac{B_{j}^{\pare{l_1}}B_{k}^{*\pare{l_1}}e^{\abs{\alpha}^2} e^{\abs{\beta}^2}}{(j-l_1)!(j-l_1-l)!} \nonumber \\
& & \hspace{10mm} \times \cor{\frac{\partial^{j-l_1}\partial^{k-l_1}}{\partial\alpha^{j-l_1}\partial\alpha^{*k-l_1}}\delta\pare{\alpha}\delta\pare{\alpha^{*}}}\cor{\frac{\partial^{j-l_1-l}\partial^{k-l_1-l}}{\partial\beta^{j-l_1-l}\partial\beta^{*k-l_1-l}}\delta\pare{\beta}\delta\pare{\beta^{*}}}.
\label{Eq:P-function}
\end{eqnarray}
Then, by substituting Eq.~\eqref{Eq:P-function} into Eq.~\eqref{Eq:Photon_dist}, we can make use of the identity
\begin{equation}
\int F\pare{\alpha,\alpha^{*}}\frac{\partial^{2\pare{j-l}}}{\partial\alpha^{j-l}\partial\alpha^{*j-l}}\delta\pare{\alpha}\delta\pare{\alpha^{*}} = \frac{\partial^{2\pare{j-l}}F\pare{\alpha,\alpha^{*}}}{\partial\alpha^{j-l}\partial\alpha^{*j-l}}\big\vert_{\alpha,\alpha^{*}=0},
\end{equation}
to obtain
\begin{eqnarray}
p\pare{n,m} &=& \frac{1}{n!m!}\sum_{j=l_1}^{\infty}\sum_{k=l_1}^{\infty}\frac{B_{j}^{\pare{l_1}}B_{k}^{*\pare{l_1}}}{\sqrt{\pare{j-l_1}!\pare{k-l_1}!\pare{j-l_1-l}!\pare{k-l_1-l}!}} \nonumber \\
& & \hspace{5mm} \times \frac{\partial^{j-l_1}\partial^{k-l_1}}{\partial\alpha^{j-l_1}\partial\alpha^{*k-l_1}}\llav{\pare{\eta\alpha^{*}\alpha + \nu}^{n}e^{-\cor{\pare{\eta -1}\alpha^{*}\alpha + \nu}}}\Big\vert_{\alpha,\alpha^{*}=0} \nonumber \\
& & \hspace{10mm} \times \frac{\partial^{j-l_1-l}\partial^{k-l_1-l}}{\partial\beta^{j-l_1-l}\partial\beta^{*k-l_1-l}}\llav{\pare{\eta\beta^{*}\beta + \nu}^{m} e^{-\cor{\pare{\eta -1}\beta^{*}\beta + \nu}}}\Big\vert_{\beta,\beta^{*}=0}.
\end{eqnarray}
Note that this photon distribution can then be used to obtain important non-classicality measures such as the Agarwal parameter and the matrix of moments of the asymmetric photon-subtracted state. Finally, notice that by setting $l=0$ in Eq. (S.9), we recover the joint photon distribution of the symmetric photon-subtraction case presented in the main text (Eq.~(1)).

\section*{\large{Matrix of moments}}
For a single PNR detector the probability generating formula is given as
\begin{align}
f(x) =\sum_{m=0}^{\infty}p(m)x^{m} =\sum_{m=0}^{\infty}\left\langle:\frac{(\eta\hat{n}+\nu)^{m}}{m!}e^{-(\eta\hat{n}+\nu)}:\right\rangle x^{m}=
\left\langle: e^{(x-1)(\eta \hat{n}+\nu)}:\right\rangle.
\label{eq:Fx}
\end{align}
Having the characteristic function we can compute the $r$-th factorial moment $\left\langle \hat{n}^{(r)}\right\rangle$ of the photon-counting statistics by taking the $r$-th derivative of $f(x)$ and evaluate it at $x=1$
\begin{align}
\left\langle \hat{n}^{(r)} \right\rangle =\sum_{m=0}^{\infty}m(m-1)...(m-r+1)p(m) =\sum_{k=0}^{l}\frac{r!}{k!(r-k)!}\nu^{r-k}(\eta \hat{n})^{k}.
\label{eq:Fx}
\end{align}
This expression allows us to compute all the moments directly from the measured photon number distributions.\\
Analogously, by using Eq.~\eqref{eq:Pmn} we obtain the two-dimensional generating function
\begin{align}
f(x,y) =\left\langle: e^{(x-1)(\eta \hat{n}_{s}+\nu)}\otimes e^{(y-1)(\eta \hat{n}_{i}+\nu)}:\right\rangle.
\label{eq:Mx}
\end{align}
And the joint moments are given as
\begin{align}
\left\langle: \hat{m}_{s}^{u}\otimes \hat{m}_{i}^{v}:\right\rangle=&\sum_{m=0}^{\infty}\sum_{n=0}^{\infty}m(m-1)...(m-u+1)\times n(n-1)...(n-v+1)p(m,n)\\
&=\left.\left\langle: \frac{d^{u}}{dx^{u}}e^{(x-1)(\eta \hat{n}_{s}+\nu)}\otimes  \frac{d^{v}}{dy^{v}}e^{(y-1)(\eta \hat{n}_{i}+\nu)}:\right\rangle\right|_{x,y=1}\\
&=\left\langle: (\eta\hat{n}_{s}+\nu)^{u}\otimes (\eta\hat{n}_{i}+\nu)^{v}:\right\rangle,
\label{eq:Mxy}
\end{align}
where we have defined the operator $\hat{m}_{s,i}=(\eta\hat{n}_{s,i}+\nu)$. Now we are in position to compute the matrix of moments
\begin{align}
M^{(n,n)}=
\begin{pmatrix}
\langle: \hat{m}_{a}^{0} \hat{m}_{b}^{0}:\rangle&  \langle: \hat{m}_{a}^{1} \hat{m}_{b}^{0}:\rangle & \langle: \hat{m}_{a}^{0} \hat{m}_{b}^{1}:\rangle & ...\\
\langle: \hat{m}_{a}^{1} \hat{m}_{b}^{0}:\rangle&  \langle: \hat{m}_{a}^{2} \hat{m}_{b}^{0}:\rangle &  \langle: \hat{m}_{a}^{1} \hat{m}_{b}^{1}:\rangle& ...\\
\langle: \hat{m}_{a}^{0} \hat{m}_{b}^{1}:\rangle&  \langle: \hat{m}_{a}^{1} \hat{m}_{b}^{1}:\rangle &  \langle: \hat{m}_{a}^{0} \hat{m}_{b}^{2}:\rangle & ...\\
\vdots&  \vdots & \vdots & \ddots
  \end{pmatrix}.
\label{eq:Mxy}
\end{align}
The principal second order minor is
\begin{align}
M^{(2,2)}=
\begin{pmatrix}
\langle: \hat{m}_{a}^{0} \hat{m}_{b}^{0}:\rangle&  \langle: \hat{m}_{a}^{1} \hat{m}_{b}^{0}:\rangle & \langle: \hat{m}_{a}^{0} \hat{m}_{b}^{1}:\rangle\\
\langle: \hat{m}_{a}^{1} \hat{m}_{b}^{0}:\rangle&  \langle: \hat{m}_{a}^{2} \hat{m}_{b}^{0}:\rangle &  \langle: \hat{m}_{a}^{1} \hat{m}_{b}^{1}:\rangle\\
\langle: \hat{m}_{a}^{0} \hat{m}_{b}^{1}:\rangle&  \langle: \hat{m}_{a}^{1} \hat{m}_{b}^{1}:\rangle &  \langle: \hat{m}_{a}^{0} \hat{m}_{b}^{2}:\rangle\\
  \end{pmatrix}.
\label{eq:Mxy}
\end{align}
And its determinant is given as
\begin{align}
det(M^{(2,2)}) =& \left\langle: \hat{m}_{s}^{2} :\right\rangle \left\langle: \hat{m}_{i}^{2} :\right\rangle
-\left\langle: \hat{m}_{s} :\right\rangle^{2} \left\langle: \hat{m}_{i}^{2} :\right\rangle
- \left\langle: \hat{m}_{s}^{2} :\right\rangle \left\langle: \hat{m}_{i} :\right\rangle^{2}
- \left\langle: \hat{m}_{s} :\right\rangle^{2} \left\langle: \hat{m}_{i} :\right\rangle^{2}\\
&+\left\langle: \hat{m}_{s} :\right\rangle^{2} \left\langle: \hat{m}_{i} :\right\rangle^{2}-\left(\left\langle: \hat{m}_{s} :\right\rangle\left\langle: \hat{m}_{i} :\right\rangle- \left\langle: \hat{m}_{s} :\right\rangle \left\langle: \hat{m}_{i} :\right\rangle\right) ^{2}\\
&=\left\langle: (\Delta\hat{m}_{s})^{2} :\right\rangle \left\langle: (\Delta\hat{m}_{i})^{2} :\right\rangle-
\left\langle: \Delta\hat{m}_{s}\Delta\hat{m}_{i} :\right\rangle^{2},
\label{eq:DetM}
\end{align}
where we have used the variances of the marginal distributions $\left\langle: (\Delta\hat{m}_{s,i})^{2} :\right\rangle=\left\langle: \hat{m}_{s,i}^{2} :\right\rangle-\left\langle: \hat{m}_{s,i} :\right\rangle^{2}$ and the covariance square $\left\langle: \Delta\hat{m}_{s}\Delta\hat{m}_{i} :\right\rangle^{2}=\left(\left\langle: \hat{m}_{s} :\right\rangle\left\langle: \hat{m}_{i} :\right\rangle-\left\langle: \hat{m}_{s} :\right\rangle\left\langle: \hat{m}_{i} :\right\rangle\right)^2$.\\
We must understand that all classically correlated light fields satisfy the inequality
\begin{align}
\left\langle: (\Delta\hat{m}_{s})^{2} :\right\rangle \left\langle: (\Delta\hat{m}_{i})^{2} :\right\rangle\geq \left\langle: \Delta\hat{m}_{s}\Delta\hat{m}_{i} :\right\rangle^{2}.
\label{eq:DetM}
\end{align}
Hence, any violation of this inequality implies that the field under study is non-classically correlated. In other words, to test the non-classicality of the correlations between the two-modes of an optical field it is sufficient to show that $\left\langle: \Delta\hat{m}_{s}\Delta\hat{m}_{i} :\right\rangle^{2}$ is greater than $\left\langle: (\Delta\hat{m}_{s})^{2} :\right\rangle \left\langle: (\Delta\hat{m}_{i})^{2} :\right\rangle$, or equivalently that the above determinant, $det(M^{(2,2)})$, is negative.

Importantly, the elements of $M^{(2,2)}$ can be computed directly from the measured joint photon number distribution
\begin{align}
\langle: \hat{m}_{s}^{1} \hat{m}_{i}^{0}:\rangle = \langle: \hat{m}_{s}^{1}:\rangle = \sum_{m=0}^{\infty}\sum_{n=0}^{\infty}mp(m,n),
\label{eq:M1}
\end{align}
\begin{align}
\langle: \hat{m}_{s}^{0} \hat{m}_{i}^{1}:\rangle = \langle: \hat{m}_{i}^{1}:\rangle = \sum_{m=0}^{\infty}\sum_{n=0}^{\infty}np(m,n),
\label{eq:M2}
\end{align}
\begin{align}
\langle: \hat{m}_{s}^{1} \hat{m}_{s}^{1}:\rangle = \langle: \hat{m}_{s}^{2}:\rangle = \sum_{m=0}^{\infty}\sum_{n=0}^{\infty}m(m-1)p(m,n),
\label{eq:M3}
\end{align}
\begin{align}
\langle: \hat{m}_{i}^{1} \hat{m}_{i}^{1}:\rangle = \langle: \hat{i}_{s}^{2}:\rangle = \sum_{m=0}^{\infty}\sum_{n=0}^{\infty}n(n-1)p(m,n),
\label{eq:M4}
\end{align}
\begin{align}
\langle: \hat{m}_{i}^{1} \hat{m}_{s}^{1}:\rangle = \sum_{m=0}^{\infty}\sum_{n=0}^{\infty}mnp(m,n).
\label{eq:M5}
\end{align}

\section*{\large{Ideal photon-subtracted two mode squeezed vacuum states}}
\begin{figure*}[t!]
\hbox{\hspace{0mm} \includegraphics[width=\textwidth]{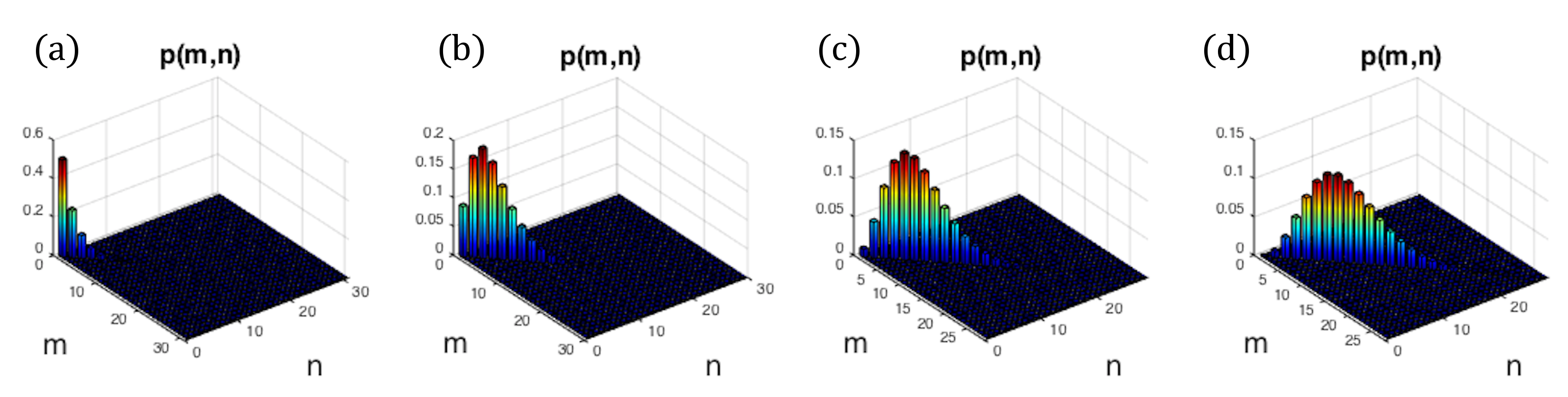}}
\caption{Theoretical joint probability distributions $p(m,n)$ for ideal photon-subtracted TMSVS using $z=0.7$. Figures (a), (b), (c), and (d) we depict $\ket{z,-l}$ for $l=0, 1, 2, 3$, respectively.}
\label{Fig:JD}
\end{figure*}

\begin{figure}[h!]
\centering
\includegraphics[width=0.75\textwidth]{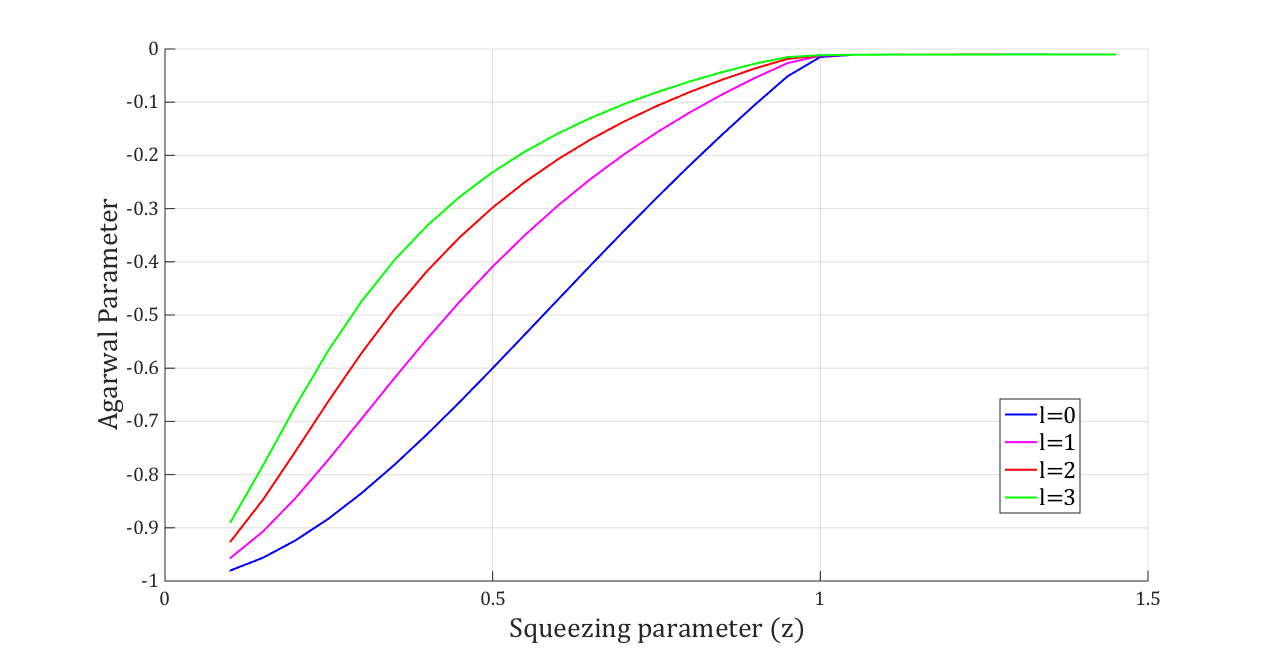}
\caption{Agarwal parameter I for ideal photon-subtracted TMSVS as a function of the squeezing parameter $z$. }
\label{Fig:AP}
\end{figure}

\begin{figure}[h!]
\centering
\includegraphics[width=0.75\textwidth]{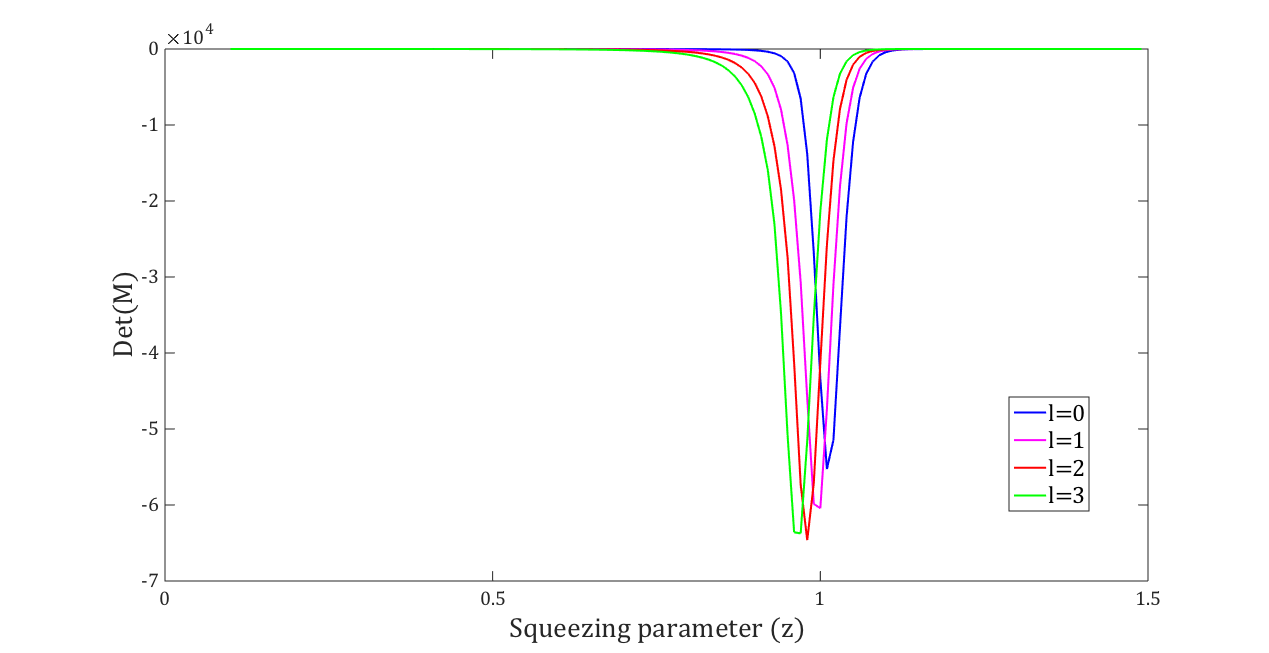}
\caption{Determinant of $M^{(2,2)}$ for $l=0,1,2,3$ as a function of $z$. }
\label{Fig:DetM}
\end{figure}

\begin{figure}[h!]
\centering
\includegraphics[width=0.75\textwidth]{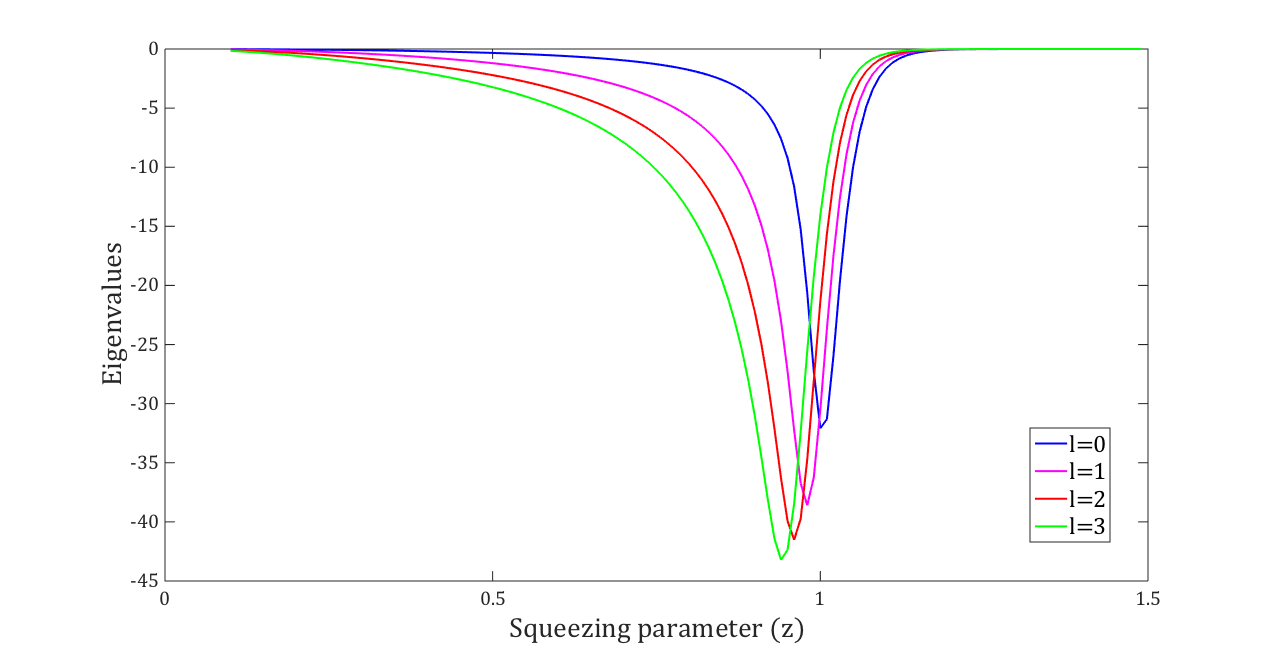}
\caption{Eigenvalues of $M^{(2,2)}$ as a function of $z$. }
\label{Fig:Ei}
\end{figure}

For the sake of completeness we briefly review the case of ideal photon-subtracted squeezed two mode vacuum states
\begin{align}
\ket{z,-l}=\sum_{j=l}^{\infty}B_{j}^{\pare{l}}\ket{j-l}_{s}\ket{j-l}_{i},
\end{align}
with $B_{j}^{\pare{l}}=\mathcal{N}^{-1/2}\cor{j!/\pare{j-l}!}z^{j}$, and the normalization constant $\mathcal{N}=\sum_{n=0}^{\infty}\abs{z}^{2j}\cor{j!/\pare{j-l}!}^2$. The joint photon number probability distribution $p(m,n)=\left|\bra{n,n}\ket{z,-l}\right|^{2}$ is shown in Fig. S\ref{Fig:JD}. As oppose to the experimental results, in the ideal case we see that $p(m,n)$ only exhibits probability terms along the diagonal.
The violation of the Cauchy-Schwarz inequality for different values of the squeezing parameter $z$ and $l=0,1,2,3$ is probed by the Agarwal parameter as shown in Fig. S\ref{Fig:AP}. For all cases, we find that $I$ is negative and it becomes less negative as we increase $z$. However, since $z\in[0,1)$, we have that $I\in[-1,0)$, and as a result, the states never become separable. As pointed out above, the nature of the correlations of these states can be analyzed through the determinant of the second order matrix of moments $M^{(2,2)}$. In Fig. S\ref{Fig:DetM} we present the determinants of $M^{(2,2)}$ as a function of $z$. These theoretical results clearly elucidate that by subtracting a higher number of photons, the square covariance $\left\langle: \Delta\hat{m}_{s}\Delta\hat{m}_{i} :\right\rangle^{2}$ becomes much bigger than the product of the marginal variances $\left\langle: (\Delta\hat{m}_{s})^{2} :\right\rangle \left\langle: (\Delta\hat{m}_{i})^{2} :\right\rangle$ and the determinant becomes more negative.\\
Finally, we compute the eigenvalues of $M^{(2,2)}$ to explore its non-negativity. To do so, we use Eqs.~(\ref{eq:M1}-\ref{eq:M5}) along with the joint photon number distributions in Fig. S\ref{Fig:AP}, and the eigenvalues are shown in Fig. S\ref{Fig:Ei}. As clearly seen, in the interval $z\in[0,1)$ all the eigenvalues are negative revealing the violation of the non-negativity of $M^{(2,2)}$, that is, $M\ngeq0$.\\

\section*{\large{2 Additional Experimental Results}}

In this section, we provide additional experimental results. The joint photon number distributions for the generated photon-subtracted TMSVS with a squeezing parameter $z$ of 0.47 and an overall system efficiency of 16.25$\%$ is shown in Fig. S4. The first row shows experimental results whereas the second row shows our theoretical predictions for our protocol. Similarly, in Fig. S5, we plot our experimental and theoretical results for a squeezing parameter $z$ of 0.1 under the same losses conditions.

\begin{figure}[h!]
\centering
\includegraphics[width=\textwidth]{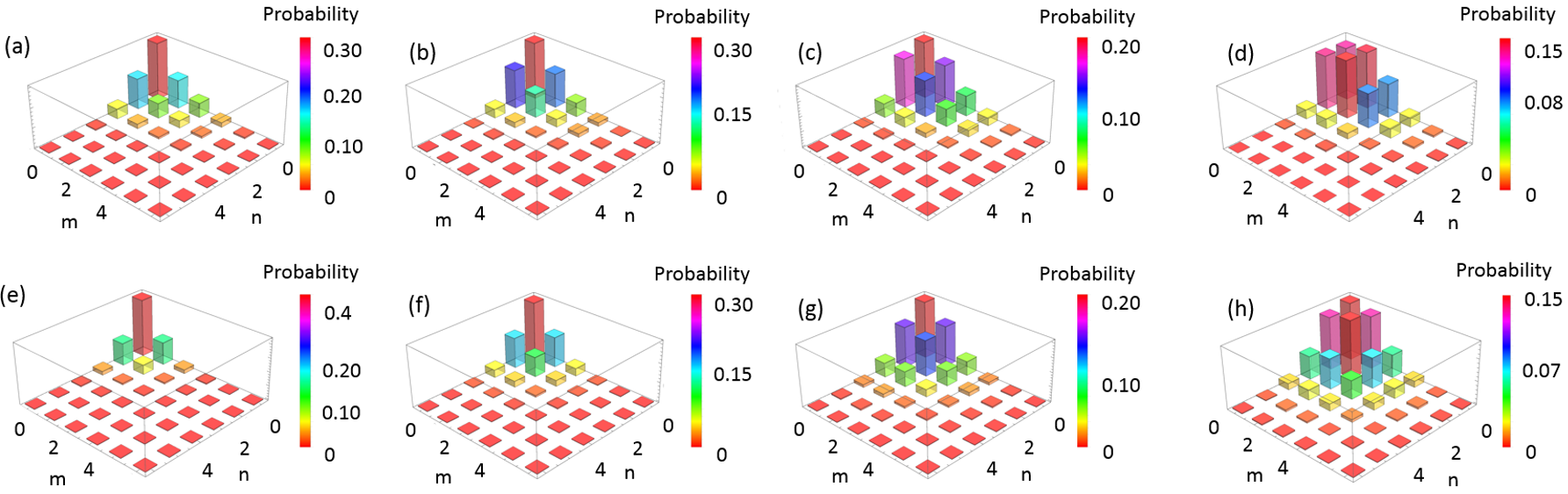}
\caption{Joint photon number distributions for  photon-subtracted TMSVS. (a to d) The first row shows the experimental joint photon number distributions for the simultaneous subtraction of zero ($l=0$), one ($l=1$), two ($l=2$) and three ($l=3$) photons. (e to h) The bottom row shows the theoretical predictions obtained by evaluating Eq. S.5. We assume an overall system efficiency of approximately $16.25\%$ and a squeezing parameter $z$ of $0.47$.}
\label{fig:Distribution3D}
\end{figure}

\begin{figure}[h!]
\centering
\includegraphics[width=12cm]{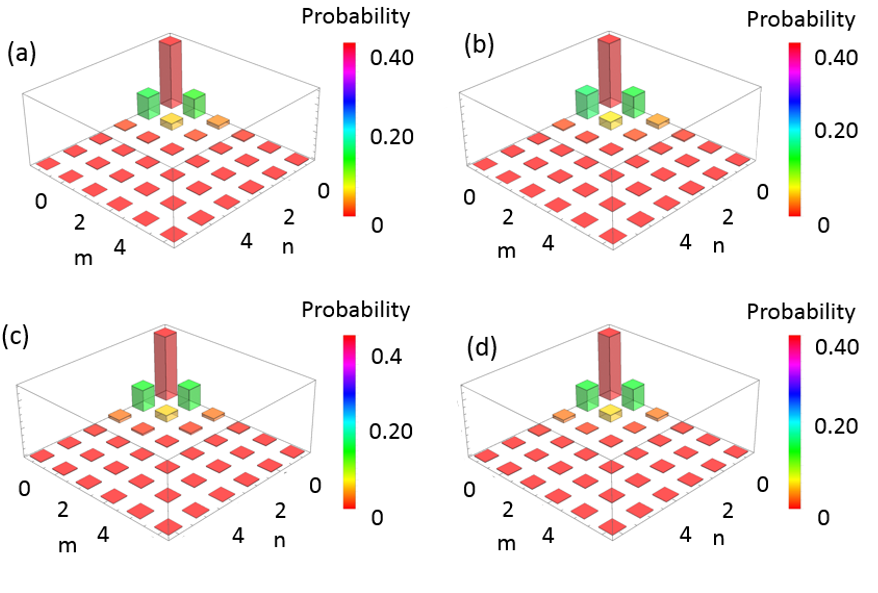}
\caption{Joint photon number distributions for  photon-subtracted TMSVS. (a to b) The first row shows the experimental joint photon number distributions for the simultaneous subtraction of zero ($l=0$), and one ($l=1$) photons. (c to d) The bottom row shows the theoretical predictions obtained by evaluating Eq. S.5. We assume an overall system efficiency of approximately $16.25\%$ and a squeezing parameter $z$ of $0.1$.}
\label{fig:Distribution3D}
\end{figure}

The results in Fig. S4 and Fig. S5 were utilized to estimate the Agarwal parameter and entropy, see Figure 4 in the main text.

\newpage
\section*{References}
\vspace{-0.5cm}
\renewcommand{\refname}{\large{}}